\documentclass[
prl,aps,
reprint,
a4paper,
superscriptaddress,
longbibliography,
floatfix
]{revtex4-1}
\usepackage{times}
\usepackage{amsfonts}
\usepackage{amsmath}
\usepackage{amssymb}
\usepackage{color}
\usepackage{multirow}
\usepackage[colorlinks=true,linkcolor=blue,citecolor=magenta,urlcolor=blue]{hyperref}

\begin{document}


\title{Comment on ``Franson Interference Generated by a Two-Level System''}


\author{Jonathan Jogenfors}
\affiliation{Institutionen f\"or systemteknik, Link\"opings Universitet, 581 83
	Link\"oping, Sweden}

\author{Ad\'an Cabello}
\affiliation{Departamento de F\'{\i}sica Aplicada II, Universidad de Sevilla,
	E-41012 Sevilla, Spain}

\author{Jan-\AA{}ke Larsson}
\affiliation{Institutionen f\"or systemteknik, Link\"opings Universitet, 581 83
	Link\"oping, Sweden}




\begin{abstract}
In a recent Letter [Phys. Rev. Lett. \textbf{118}, 030501 (2017)], Peiris,
Konthasinghe, and Muller report a Franson interferometry experiment using pairs
of photons generated from a two-level semiconductor quantum dot. The authors
report a visibility of $66\%$ and claim that this visibility ``goes beyond the
classical limit of $50\%$ and approaches the limit of violation of Bell's
inequalities ($70.7\%$).'' We explain why we do not agree with this last
statement and how to fix the problem.
\end{abstract}



\maketitle


In a recent Letter~\cite{PKM17}, Peiris, Konthasinghe, and Muller report a
Franson interferometry experiment using pairs of photons generated via
frequency-filtered scattered light from a two-level semiconductor quantum dot.
The authors report a visibility of $66\%$ and claim that this visibility ``goes
beyond the classical limit of $50\%$ and approaches the limit of violation of
Bell's inequalities ($70.7\%$).'' In the following we explain why we do not
agree with this last statement.

A violation of the Clauser-Horne-Shimony-Holt (CHSH) Bell inequality
\cite{CHSH69} without supplementary assumptions (so that it is loophole-free
and therefore potentially usable for device-independent applications)  using a
maximally entangled state is only possible in a very small region of values of
the overall detection efficiency $\eta$ and the visibility $V$. Specifically, it
must occur that $V \ge (2/\eta-1)/\sqrt{2}$ \cite{Larsson99}. Therefore, the
70.7\% visibility bound mentioned by Peiris, Konthasinghe, and Muller only holds
under the assumption that $\eta =1$.

The problem is that this value is impossible to achieve in the Franson
interferometer, even ideally. As the authors correctly point out, in the
Franson interferometer there is a crucial postselection step which requires
discarding, on average, 50\% of the recorded photons. Therefore, {\em even in
the ideal case that the detectors and couplings were perfect}, the effective
$\eta$ falls to $50\%$. This implies that it is possible to produce a classical
local hidden variable models while retaining the same output statistics as
predicted by quantum theory~\cite{AKLZ99,CRVDM09,JL14}.

In fact, the above problem has recently been exploited to experimentally show
that the security proof in Franson-based quantum key distribution schemes can
be circumvented, exposing its users to eavesdropping~\cite{JEABL15}. In these
attacks, tailored pulses of classical light are used, which indicates that the
$50\%$ ``classical limit'' can be beat even in a purely classical setting.

However, as described in \cite{AKLZ99}, there is a possibility of detecting a
genuine violation of a Bell inequality in the setting of Peiris, Konthasinghe,
and Muller. It requires using a {\em different} Bell inequality, namely, a
three-setting chained Bell inequality introduced by Pearle \cite{Pearle70}.
This modification allows for a genuine violation of local realism, but requires
a higher visibility: At least, $94.63\%$ \cite{AKLZ99,JL14}. Although
demanding, a recent work \cite{TMJVLV17} shows that such an experiment is
feasible.

In conclusion, while the setup in \cite{PKM17} is promising, the experimental
data does not rule out all classical descriptions. A test of the three-setting
chained Bell inequality could be a more suitable application for this
correlated photon pair source. However, the corresponding experiment would be
much more challenging as it requires a visibility of, at least, $94.63\%$.


\begin{acknowledgments}
This work was supported by the project ``Photonic Quantum Information'' (Knut
and Alice Wallenberg Foundation, Sweden) and Project No.\ FIS2014-60843-P,
``Advanced Quantum Information'' (MINECO, Spain), with FEDER funds.
\end{acknowledgments}




\end{document}